\documentclass[journal]{IEEEtran}
\usepackage{algorithmic}
\usepackage{algorithm}
\usepackage{multirow}
\usepackage{graphicx}
\usepackage{mathtools, cuted}
\usepackage{times}
\usepackage{setspace,amsmath,latexsym,cite,amssymb,epsfig,amsfonts}
\usepackage{color}
\usepackage{url}
\usepackage{epstopdf}
\usepackage{psfrag}
\usepackage{footmisc}
\usepackage{caption}
\usepackage{subcaption}
\usepackage{authblk}
\usepackage{multicol}
\usepackage{algorithmic}
\usepackage{algorithm}
\usepackage{stfloats}
\ifCLASSINFOpdf

\else

\fi
\begin{document}
	\title{Impact of Phase-Noise and Spatial Correlation on Double-RIS-Assisted Multiuser MISO Networks}
	
	\author{Zaid~Abdullah,~\IEEEmembership{Member,~IEEE,}
		Anastasios~Papazafeiropoulos,~\IEEEmembership{Senior Member,~IEEE,}
		Steven~Kisseleff,~\IEEEmembership{Member,~IEEE,}
		Symeon~Chatzinotas,~\IEEEmembership{Senior Member,~IEEE,}
		and~Bj$\ddot{\text{o}}$rn~Ottersten,~\IEEEmembership{Fellow,~IEEE}
		\thanks{This work was supported by the Luxembourg National Research Fund (FNR) through the CORE Project under Grant RISOTTI C20/IS/14773976.
			\par Z. Abdullah, S. Kisseleff, S. Chatzinotas, and B. Ottersten  are with the Interdisciplinary Centre for Security, Reliability and Trust (SnT), University of Luxembourg, L-1855, Luxembourg. (Emails: \{zaid.abdullah, steven.kisseleff, symeon.chatzinotas, bjorn.ottersten\}@uni.lu).
			\par Anastasios Papazafeiropoulos is with the Communications and Intelligent
			Systems Research Group, University of Hertfordshire, Hatfield AL10 9AB,
			U.K., and also with the SnT, University of Luxembourg, Luxembourg. Email: tapapazaf@gmail.com.}}
	
	\maketitle
	
	\begin{abstract}
		We study the performance of a phase-noise impaired double reconfigurable intelligent surface (RIS)-aided multiuser (MU) multiple-input single-output (MISO) system under spatial correlation at both RISs and base-station (BS). The downlink achievable rate is derived in closed-form under maximum ratio transmission (MRT) precoding. In addition, we obtain the optimal phase-shift design at both RISs in closed-form for the considered channel and phase-noise models. Numerical results validate the analytical expressions, and highlight the effects of different system parameters on the achievable rate. Our analysis shows that phase-noise can severely degrade the performance when users do not have direct links to both RISs, and can only be served via the double-reflection link. Also, we show that high spatial correlation at RISs is essential for high achievable rates.   
	\end{abstract}
	
	\IEEEpeerreviewmaketitle

	\section{Introduction}
	\IEEEPARstart{T}{oday}, there is no lack of interest among researchers around the globe when it comes to the potential of reconfigurable intelligent surfaces (RISs) \cite{pan2021reconfigurable}. This does not come as a surprise since the concept of RIS shattered the old belief that the response of a wireless channel cannot be altered. Also, a great benefit of this technology lies in the fact that RISs are considered as energy green nodes since no (or limited number of) radio-frequency chains are required for their operations, and no power amplification is involved \cite{wu2019towards}. 
	\par Despite the large volume of existing research on RISs, most of the works considered a single RIS-assisted network. However, in practical scenarios, signals can be subject to reflections from multiple spatially-separated RISs \cite{huang2021multi}. To that end, the authors in \cite{zheng2021double} studied a cooperative beamforming design of a multiuser (MU) system based on instantaneous channel state information (iCSI), and the estimation process of such iCSI for different communication links in the double-RIS networks was demonstrated in \cite{you2020wireless}.  The authors in \cite{papa2} derived the coverage probability of a single-input single-output (SISO) double-RIS network with statistical CSI (sCSI)-based reflect beamforming design (RBD), where the optimization of phase-shifts at RISs is performed without requiring any knowledge of iCSI. Note that such approach is very useful in practice as the amount of training required to estimate all different channels in multiple RISs networks can become very large with respect to the coherence interval of a wireless channel. 
	\par Another practical aspect is the hardware impairments (HWIs), which cannot be avoided in realizing any communication system. Specifically, two main types of HWIs have been highlighted in the literature regarding the implementation of RISs. The first one is the discrete phase-shifts \cite{wu2019towards}, where only few quantization levels of reflecting coefficients can be obtained; while the second type of HWIs is the phase-noise, which reflects imperfections in phase-estimation and/or phase-quantization \cite{badiu2019communication}. The latter will be the focus of this work. \par Few works have investigated the phase-noise in RIS-assisted networks so far. For example, the authors in \cite{badiu2019communication} and \cite{wang2021outage} thoroughly analyzed the performance of a point-to-point SISO system assisted by a RIS impaired with phase-noise under fading and non-fading (i.e. line-of-sight) channels, respectively. In addition, the works in \cite{papa1} and \cite{papa3} studied the performance of an MU multiple-input single-output (MISO) system assisted by a single RIS under HWIs, where the gradient ascent method was adopted for the RIS RBD. In \cite{xing2021achievable} the authors investigated the achievable rate of a RIS assisted SISO network with HWIs, where the RBD problem was tackled via the semidefinite relaxation approach. Finally, the work in \cite{zhou2020spectral} investigated the spectral and energy efficiencies of a single-user MISO network aided by an imperfect RIS, and their results showed that HWIs limit the spectral efficiency even when the number of reflecting elements grows to infinity.  
	\par All previous works considered the impact of HWIs on a single-RIS network. In contrast, here we analyze the achievable rate of a double-RIS assisted MU-MISO network under phase-noise impacting both RISs. In addition, we obtain a closed-form solution for the optimal RBD and shed light into the role of spatial correlation at both the base-station (BS) and RISs, and their effect on the achievable rate. Our contributions are summarized as follows: 
	\begin{itemize}
		\item We derive a closed-form expression for the achievable rate of correlated and phase-noise impaired double-RIS MU-MISO network under maximum ratio transmission (MRT) precoding at BS and optimized RBD at both RISs.
		\item We demonstrate that the spatial correlation at  RISs leads to higher channels gain, thereby resulting in an improved rate performance, while the transmit correlation at BS has a negative impact on the achievable rates. 
		\item In addition, we prove that based on sCSI, the phase-shift matrices at both RISs that lead to a maximum achievable rate can be obtained in closed-form without the need for any sophisticated optimization tools.  
	\end{itemize}To the best of our knowledge, these aspects have not been investigated so far in the literature. 
	\par The rest of this paper is organized as follows: Section \ref{system_model} presents the system model. Section \ref{analysis} analyzes the achievable rate and RBD. Section \ref{results} presents the numerical results and their discussion. Conclusions are drawn in Section \ref{conclusions}.
	\par \textit{Notations}: Matrices and vectors are denoted by boldface uppercase and lowercase letters, respectively.	$\boldsymbol a^T$, $\boldsymbol a^H$, $\boldsymbol a^\ast$, and $\left\| \boldsymbol a \right\|$ are the transpose, Hermitian transpose, conjugate, and Euclidean norm of a vector $\boldsymbol a$, respectively. $\boldsymbol I_N$ is the $N\times N$ identity matrix, while $[\boldsymbol A]_{i,j}$ is the $(i,j)$th entry of $\boldsymbol A$, and $[\boldsymbol a]_i$ is the $i$th element of $\boldsymbol a$. The absolute, expected, and trace operators are denoted by $\left|\cdot \right|$, $\mathbb E\{\cdot{}\}$, and $\text{tr}(\cdot)$, respectively. Furthermore, $\mathrm {diag} \{\boldsymbol a\}$ is a diagonal matrix whose diagonal contains the elements of $\boldsymbol a$, while $\mathrm {diag}\{\boldsymbol A\}$ is a vector whose elements are the diagonal of $\boldsymbol A$. Finally, $\Re{\{x\}}$ denotes the real part of a complex number $x$.
	\begin{figure}[t]
		\centering
		{\includegraphics[width=7cm,height=5cm, trim={0cm 0cm 0cm 0cm},clip]{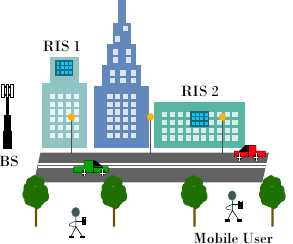}}
		\caption{The considered double RIS communication system in a scattering environment.}
		\label{system}
	\end{figure}
	\section{System Model}\label{system_model}
	\subsection{Signal, channel, and phase-noise models}
	We consider a time-division-duplex MU downlink scenario where there is one BS consisting of a uniform linear array (ULA) with $M$ active radiating elements, communicating with $K$ single-antenna users $\{U_1, \cdots, U_K\}$. We focus on a challenging, but meaningful scenario that justifies the presence of RISs where the direct path between the BS and all $K$ users is assumed to be blocked, and communication is facilitated via two  RISs ($I_1$ and $I_2$) equipped with $N_1$ and $N_2$ reflecting elements, respectively, as shown in Fig. \ref{system}. A rectangular geometry is adopted for each RIS with $N_{V_i}$ vertical and $N_{H_i}$ horizontal elements, such that $N_i = N_{V_i}N_{H_i}$, $i\in\{1,2\}$. Assuming that all communication links experience flat fading, the received signal at $k$th user is
	{\small \begin{align}\label{eq1} 
			y_k = \boldsymbol h_k^T\boldsymbol w_k x_k + \sum_{l\in \mathcal U\setminus k} \boldsymbol h_k^T\boldsymbol w_l x_l + z_k,
	\end{align}}where $\mathcal U = \{1, \cdots, K\}$, $x_k$ is the information symbol intended for $k$th user satisfying $\mathbb E\{|x_k|^2\}=p_k$ with $p_k$ being the power allocated for $k$th user, such that $\sum_{k\in \mathcal U} p_k \le P_t$, where $P_t$ is the total power available at BS. Also, $z_k \sim \mathcal{CN}(0, \sigma^2)$ is the additive white Gaussian noise (AWGN) at $U_k$, $\small {\boldsymbol w_k = \frac{\boldsymbol h^\ast_k}{\sqrt{\mathbb E\{\left \| \boldsymbol h_k \right \|^2\}}} \in \mathbb C^{M\times 1}}$ is the MRT beamforming vector at BS for $U_k$, and $\boldsymbol h_k  \in \mathbb C^{M\times 1}$ is the effective channel vector for the same user given as{\footnote{To focus on the effects of phase-noise and spatial correlation, we
		assume perfect knowledge of CSI for the overall effective channel vector $\boldsymbol h_k$, while it can be obtained via traditional estimation techniques in practice \cite{bjornson2017massive}. Thus, our results represent an upper-bound for imperfect CSI scenarios.}}
	{\small \begin{align} \label{h_k}
			\boldsymbol h_k = \boldsymbol{H}_{B1} \tilde{\boldsymbol{\Theta}}_{1}\boldsymbol{G} \tilde{\boldsymbol{\Theta}}_{2}\boldsymbol{q}_{2k}  + \boldsymbol{H}_{B1} \tilde{\boldsymbol{\Theta}}_{1} \boldsymbol{q}_{1k} + \boldsymbol{H}_{B2} \tilde{\boldsymbol{\Theta}}_{2} \boldsymbol{q}_{2k},
	\end{align}}where the first term in (\ref{h_k}) accounts for double reflection link,\footnote{There exists a path from BS $\rightarrow I_2 \rightarrow I_1 \rightarrow U_k$. However, its gain is sufficiently weaker than that from BS $\rightarrow I_1\rightarrow I_2 \rightarrow U_k$ shown in (\ref{h_k}), since $I_1$ is closer to BS than $I_2$, and thus the former path can be neglected.} while the second and third terms account for single reflection links through the first and second RISs, respectively. In particular, $\boldsymbol q_{ik}$ is the vector of channel coefficients between $I_i\rightarrow U_k$, such that $\boldsymbol q_{ik} = \sqrt{\beta_{ik}}\boldsymbol R_{i}^{\frac{1}{2}} \tilde{\boldsymbol q}_{ik} \in \mathbb C^{N_i\times 1}$, with $\tilde{\boldsymbol q}_{ik} \sim \mathcal{CN}(\boldsymbol 0, \boldsymbol I_{N_i}) \in \mathbb C^{N_i\times 1}$ being a vector containing independent and identically distributed (i.i.d.) Rayleigh fading components, while $ \boldsymbol R_i \in \mathbb C^{N_i\times N_i}$ is the Hermitian positive semidefinite correlation matrix for the $i$th RIS with $\text{tr}(\boldsymbol R_i) = N_i$, and $\beta_{ik}$ is the channel gain between each reflecting element at $I_i$ and $U_k$ given as $\beta_{ik} = d_{ik}^{-\alpha}A$ \cite{bjornson2020rayleigh}, where $A = d_V d_H$ is the area of each reflecting element with $d_V$ and $d_H$ being the vertical height and horizontal width, $d_{ik}$ is the distance between $I_i$ and $U_k$, and $\alpha$ is the path-loss exponent. On the other hand, $\boldsymbol H_{Bi} \in\mathbb C^{M\times N_i}$ is a matrix containing channel coefficients between BS and $I_i$. Specifically, we have $\boldsymbol H_{Bi} = \sqrt{\beta_{Bi}} \boldsymbol R_{B}^{\frac{1}{2}} \tilde{\boldsymbol H}_{Bi} \boldsymbol R_i^{\frac{1}{2}}$, where $\beta_{Bi} = d_{Bi}^{-\alpha} A$ is the channel gain between BS and each reflecting element at $I_i$, $\boldsymbol R_B\in\mathbb C^{M\times M}$ is a Hermitian positive semidefinite matrix that accounts for the antenna correlation at the BS with $\text{tr}(\boldsymbol R_B) = M$, and $\text{vec}\big(\tilde{\boldsymbol H}_{Bi}\big)\sim\mathcal{CN}\left(\boldsymbol 0, \boldsymbol I_{MN_i}\right)$ is the i.i.d Rayleigh fading components. Similarly, we have $\boldsymbol G = \sqrt{\beta_G}\boldsymbol R_1^{\frac{1}{2}} \tilde{\boldsymbol G} \boldsymbol R_2^{\frac{1}{2}} \in \mathbb C^{N_1\times N_2}$ contains the correlated Rayleigh fading channel coefficients between the two RISs with $\text{vec}(\tilde{\boldsymbol G}) \sim \mathcal {CN}(\boldsymbol 0, \boldsymbol I_{N_1N_2})$, and $\beta_G = d_{1,2}^{-\alpha}A^2$ is the corresponding channel gain with $d_{1,2}$ being the distance between the two RISs. \par Moreover, for the phase-shift matrix of $i$th RIS ($i\in\{1,2\}$), we have $\tilde{\boldsymbol{\Theta}}_{i} = \bar{\boldsymbol{\Theta}}_{i} \boldsymbol{\Theta}_{i} \in \mathbb C^{N_i\times N_i}$, whereas $\small{\boldsymbol{\Theta}_i = \text{diag}\left\{\left[e^{\jmath [\boldsymbol \theta_i]_1}, e^{\jmath [\boldsymbol \theta_i]_2}, \cdots, e^{\jmath [\boldsymbol \theta_i]_{N_i}} \right]\right\}}$ is the diagonal phase-shift matrix, while $\small{\bar{\boldsymbol{\Theta}}_{i} = \text{diag}\left\{\left[e^{\jmath [\bar{\boldsymbol \theta}_i]_{1}}, e^{\jmath [\bar{\boldsymbol \theta}_i]_{2}}, \cdots, e^{\jmath [\bar{\boldsymbol \theta}_i]_{N_i}} \right]\right\}}$ accounts for the phase-noise errors at the $i$th RIS. Similar to \cite{papa1,  badiu2019communication, wang2021outage}, the phase-noise, captured by $\bar{\boldsymbol \Theta}_1$ and $\bar{\boldsymbol \Theta}_2$, can be modeled according to the Von Mises (VM) distribution with zero-mean and a characteristic function (CF) $\small {\mathbb E \left\{e^{\jmath [\bar{\boldsymbol \theta}_i]_{n}} \right\} = \frac{I_1(\kappa)}{I_0(\kappa)} = \varphi}$, where $I_p (\kappa)$ is the modified Bessel function of first kind and order $p$, and $\kappa$ is the concentration parameter accounting for the estimation accuracy.
	\subsection{Spatial correlation models}
	Here, we introduce the adopted spatial correlation models for BS and both RISs.
	\subsubsection{Spatial correlation at BS} We apply the Kronecker correlation model \cite{965098} at the BS, which is suitable for uniform linear arrays, such that the $(i,j)$th entry of $\boldsymbol R_{B}$ is given as
	\begin{equation}
		\small 
		[\boldsymbol R_B]_{i,j} = 	\begin{cases}
			\rho^{(j-i)}, & \text{if}\ i\le j\\
			(\rho^{|j-i|})^\ast, & \text{otherwise},
		\end{cases} \ \ \ \ \ \ \forall \{i,j\} \in \mathcal M,
	\end{equation}where $\rho \in \mathbb C$ is the correlation coefficient satisfying $|\rho|\le 1$, and $\mathcal M = \{1, 2, \cdots, M\}$.
	\subsubsection{Spatial correlation at RISs}For each RIS, we utilize the spatial correlation proposed by \cite{bjornson2020rayleigh} for rectangular surfaces. In particular, the ($l,m$)th entry of $\boldsymbol R_i$ is given as
	\begin{equation} \label{R_i} 
		\small 
		\left[\boldsymbol R_i\right]_{l,m} = \mathrm{sinc} \left( \frac{2\left\|\boldsymbol u_{i_l} - \boldsymbol u_{i_m} \right\|}{\lambda}\right), \ \ \ \ \ \ \forall \{l,m\} \in \mathcal N_i,
	\end{equation}where $\left\|\boldsymbol u_{i_l} - \boldsymbol u_{i_m} \right\|$ is the distance between $l$th and $m$th reflecting elements at $I_i$,\footnote{We denote the distance between two adjacent elements at the RIS as $\varepsilon$.} $\lambda$ is the wavelength, and $\mathcal N_i = \{1, 2, \cdots, N_i\}$.
	\section{Achievable rate analysis and optimization} \label{analysis} 
	Here, we formulate the received signal-to-interference-plus noise ratios (SINRs) and provide closed-form expressions for the achievable rate and optimal RBD. 
	\subsection{Received SINR and achievable rate}
	Given that the instantaneous received SINR for $k$th user is  
	{\small \begin{align} \label{SINR} 
			\mathrm{SINR}_k  = \frac{p_k \left|\boldsymbol h_k^T\boldsymbol w_k\right|^2}{\sum\limits_{\substack{l\in \mathcal U \setminus k}} p_l \left|\boldsymbol h_k^T \boldsymbol w_l\right|^2+ \sigma^2 },
	\end{align}}the ergodic rate for the same user can be expressed as $\mathcal R_k = \mathbb E\{\log_2 (1+ \mathrm{SINR}_k)\}$. However, since obtaining a closed-form expression for $\mathcal R_k$ is mathematically challenging, we follow a similar approach to that in \cite{bjornson2017massive} which takes advantage of channel hardening, and derive the lower-bound achievable rate as demonstrated in the following Theorem. \\ \textbf{Theorem 1}: \textit{For a given $\boldsymbol{\Theta}_1$ and $\boldsymbol{\Theta}_2$, the downlink achievable rate for $U_k$ with MRT precoding and phase-noise is} 
	{\small \begin{align} \label{lower_bound_capacity}
			\underline{\mathcal R}_k = \log_2 \left(1+\underline{\mathrm{SINR}}_k \right)
	\end{align}}with
	{\small \begin{align} \label{lower_sinr}
			\underline{\mathrm{SINR}}_k = \frac{p_k M^2}{\text{tr} \left(\boldsymbol R_B^2\right) \Big(\sum\limits_{\substack{l\in \mathcal U}} p_l \Big) + \frac{M\sigma^2}{\eta_k}},		
	\end{align}}where $\eta_k$ is given in a closed-form as
	\small {\begin{align}\label{eta_k}
			&\eta_k  = \beta_{B1}\beta_{2k} \beta_G \Big[\text{tr}\left(\boldsymbol R_2 \bar{\boldsymbol R}_2\right) \Big( \varphi^4 \text{tr}\left(\boldsymbol R_1 \bar{\boldsymbol R}_1\right) + (\varphi^2-\varphi^4)N_1 \Big) \nonumber \\ & \hspace{2.2cm} + N_2\Big((\varphi^2-\varphi^4)\text{tr}\left(\boldsymbol R_1 \bar{\boldsymbol R}_1\right) + (1-\varphi^2)^2N_1\Big)\Big]\nonumber \\ & \hspace{0.45cm} + \sum_{i=1}^{2}\beta_{Bi}\beta_{ik} \Big(\varphi^2\text{tr}\left(\boldsymbol R_i \bar{\boldsymbol R}_i\right) + (1-\varphi^2)N_i\Big) 
	\end{align}}with $\bar{\boldsymbol R}_i = \boldsymbol \Theta_i \boldsymbol R_i {\boldsymbol{\Theta}}_{i}^H$. \par  \textit{Proof}: See the Appendix. 
	\\ \textbf{Remark 1:} The expression of SINR in (\ref{lower_sinr}) demonstrates that while spatial correlation at the BS is harmful,\footnote{Here, we consider a transmit correlation model, which reflects the spatial correlation at the BS due to limited spacing among different antenna elements and/or insufficient scattering, and such correlation is independent of the user. However, local scattering-based channel correlation where each user experience different spatial correlation with BS can in fact lead to an enhanced rate performance \cite{bjornson2017massive}.} the correlation at both RISs can in fact be utilized to enhance the achievable SINR by maximizing $\eta_k$. In fact, we will prove in the following subsection that under optimal RBD the achievable rate increases with the correlation level at each RIS, as higher correlation can always lead to increased values of $\eta_k$.
	\\ \textbf{Remark 2:} When $\varphi = 0$ (which corresponds to a VM noise concentration parameter of $\kappa = 0$), the optimization of $\boldsymbol \Theta_1$ and $\boldsymbol \Theta_2$ would not lead to any performance enhancement, and both RISs can only reflect the impinging signal without any phase-adjustment capabilities. This can be further illustrated by the probability density function formula of the VM noise given as $f{(r|\kappa)} = \frac{e^{\kappa \cos(r)}}{2\pi I_0(\kappa)}$ with $r\in[-\pi,\pi]$ \cite{wang2021outage}. Clearly, when $\kappa = 0$, the VM phase-noise will be evenly distributed over the period $[-\pi, \pi]$, and hence, no useful information regarding the phase-estimation can be obtained to facilitate the RBD. This is in line with the findings of \cite{wang2021outage} and  \cite{papa1} for the single RIS case. However, for the double RIS, we can still benefit from the additional $\beta_{B1}\beta_{2k}\beta_{G} N_1 N_2 + \beta_{B2} \beta_{2k} N_2$ channel gain compared to the single RIS case where only $I_1$ exists (in such case the channel gain would be  $\beta_{B1} \beta_{1k} N_1$), as demonstrated by (\ref{eta_k}).
	\par Finally, it is also observed from (\ref{eta_k}) that the double reflection link is the most affected by the phase-noise in terms of $\varphi^4$, compared to $\varphi^2$ for single reflection links.\footnote{Note that $0\le \varphi\le 1$ with $\varphi=1$ corresponding to an ideal case without any phase-noise.} Thus, for scenarios where users can only be served by the double-reflection link, the performance of multi-RIS networks can be severely degraded by phase-noise errors. 
	
	\par We next shift our attention to the phase-shift design of both RISs.
	\subsection{Phase-shift design}
	Our aim is to optimize $\boldsymbol \Theta_1$ and $\boldsymbol \Theta_2$, based on only sCSI,\footnote{Utilizing only sCSI to perform the RBD is highly efficient in terms of the amount of training required as only the overall cascaded channel needs to be estimated, while it leads to sub-optimal performance in terms of received SINRs compared to iCSI-based RBD. However, the latter approach requires accurate CSI knowledge of all channel links involved, which is costly in terms of the amount of training required, especially when dealing with large RISs.} to maximize the achievable rate in (\ref{lower_bound_capacity}) for all users. We start by formulating the problem as
	\begin{subequations} \label{OP1}
		\small
		\begin{align}
			& \hspace{0.5cm}\underset{\boldsymbol \Theta_1,\ \boldsymbol \Theta_2}{\text{maximize}} \hspace{0.6cm}  \sum_{k\in \mathcal U} \underline{\mathcal R}_k      \hspace{5cm} (\ref{OP1}) \nonumber \\
			&\hspace{0.5cm}\text{subject to} \hspace{0.6cm} \big|[\boldsymbol \Theta_i]_{n,n}\big| =1, \ \ \ \ \forall n\in\mathcal N_i, \ \ \ i\in\{1,2\}. \label{2b} 
		\end{align}
	\end{subequations}Interestingly, from (\ref{lower_sinr}) we can observe that the only term related to the phase-shift matrices is $\eta_k$. Thus, in order to maximize the achievable rate, we only need to maximize $\eta_k\ (\forall k\in\mathcal U)$. Now, we can further introduce the following Corollary. 
	\\ \textbf{Corollary 1}: \textit{The maximization of the achievable rate is equivalent to the maximization of} $\text{tr} \left(\boldsymbol R_i \boldsymbol \Theta_i \boldsymbol R_i \boldsymbol \Theta_i^H\right)$ for both $i=\{1,2\}$.\par \textit{Proof}: Given that the characteristic function of the VM noise satisfies $0\le \varphi\le 1$, it holds that $\varphi^p \ge \varphi^n$ for all $p\le n$, and $\{p,n\}$ are positive integers. Thus, maximizing $\eta_k$ is equivalent to maximizing the term $u_k = c_{0k} v_2v_1 + c_{1k}v_1 + c_{2k} v_2+ c_{3k}$, with the optimization variables $v_1$ and $v_2$ being $\text{tr} \left(\boldsymbol R_1 \boldsymbol \Theta_1 \boldsymbol R_1 \boldsymbol \Theta_1^H\right)$ and $\text{tr} \left(\boldsymbol R_2 \boldsymbol \Theta_2 \boldsymbol R_2 \boldsymbol \Theta_2^H\right)$, respectively, and $\{c_{0k}, \cdots, c_{3k}\}$ are constants belonging to the set of positive real numbers $\mathbb R_{++}$. \par Clearly, $u_k$ is maximum if and only if both $v_1$ and $v_2$ are maximum, which indicates that both variables should belong to $\mathbb R_{++}$. Moreover, since maximizing $v_1$ does not depend on $v_2$, and vice-versa, optimizing each of $v_1$ and $v_2$ separately would also result in both variables being jointly optimal in terms of their sum and product.  Therefore, and given that $v_1$ and $v_2$ are independent of the user index, it holds that maximizing both variables separately, via phase-optimization, will result in the optimal achievable sum rate. This concludes our proof. \par Interestingly, Corollary 1 shows that the phase optimization of different RISs in a multiple RIS-assisted scenario can be performed separately under sCSI, and without any loss of optimality. Next, for optimal RBD, we introduce the following Theorem.   
	\par \textbf{Theorem 2}:  \textit{The optimal solution for $\boldsymbol \Theta_i \ (i\in\{1,2\})$ that maximizes} $\text{tr} \left(\boldsymbol R_i \boldsymbol \Theta_i \boldsymbol R_i \boldsymbol \Theta_i^H\right)$, \textit{and hence the achievable sum rate, must satisfy $\boldsymbol \Theta_i^\star = \text{diag}\{\exp(\jmath c\boldsymbol 1_{N_i})\}$, where $c\in \mathbb R$ is any real number, and $\boldsymbol 1_{N_i}$ is a vector of length $N_i$ with entries of all ones}.
	\par \textit{Proof}:  By examining the structure of $v_i$ = $\small{\text{tr}\left( \boldsymbol R_{i} \boldsymbol{\Theta}_i \boldsymbol R_i \boldsymbol \Theta_i^H\right)}$, and keeping in mind that $\boldsymbol R_i$ is a Hermitian matrix, and the trace of a square matrix is the sum of its diagonal, we can express $v_i$ as
	{\small \begin{align}
			v_i = \sum_{n = 1}^{N_i} \sum_{l=1}^{N_i} \left|\left[\boldsymbol R_i\right]_{n,l}\right|^2 e^{\jmath \left([\boldsymbol \theta_i]_l - [\boldsymbol \theta_i]_n\right)}.
	\end{align}}Clearly, $v_i$ is maximum only when $\small{[\boldsymbol \theta_i^\star]_l = [\boldsymbol \theta_i^\star]_n}$, which means that all phase-shifts at each RIS should have equal values, i.e. $(\small{[\boldsymbol \theta_i^\star]_1 = [\boldsymbol \theta_i^\star]_2 = \cdots \cdots = [\boldsymbol \theta_i^\star]_{N_i}})$. 
	\par Theorem 2 means that for correlated Rayleigh fading, there are infinite solutions of optimal phase-shifts, and each solution can be easily generated without any sophisticated optimization tools. It follows that under optimal RBD, the expression for $\eta_k$ corresponding to the optimal achievable rate becomes 
	\small {\begin{align}\label{eta_k_optimal}
			&\eta_k^\star  = \beta_{B1}\beta_{2k} \beta_G \Big[\text{tr}\left(\boldsymbol R_2^2\right) \Big( \varphi^4 \text{tr}\left(\boldsymbol R_1^2\right) + (\varphi^2-\varphi^4)N_1 \Big) \nonumber \\ & \hspace{2.2cm} + N_2\Big((\varphi^2-\varphi^4)\text{tr}\left(\boldsymbol R_1^2 \right) + (1-\varphi^2)^2N_1\Big)\Big]\nonumber \\ & \hspace{0.45cm} + \sum_{i=1}^{2}\beta_{Bi}\beta_{ik} \Big(\varphi^2\text{tr}\left(\boldsymbol R_i^2 \right) + (1-\varphi^2)N_i\Big).
	\end{align}}\textbf{Remark 3:} Noting that the values in the off-diagonal of $\boldsymbol R_i$ reflect the amount of correlation between two reflecting elements at $I_i$, such that the higher the correlation the larger these values become,  (\ref{eta_k_optimal}) clearly demonstrates that a  higher spatial correlation at each RIS results in higher $\eta_k^\star$ (and thus higher achievable rate), since $\text{tr}(\boldsymbol R_i^2)$ is the sum of all elements in $\boldsymbol R_i$ squared, which proves our previous statement in Remark 1. Considering two extreme cases, if no correlation existed at $\boldsymbol R_i$ (i.e. $\boldsymbol R_i = \boldsymbol I_{N_i}$), then $\text{tr}(\boldsymbol R_i^2)$ would be equal to $N_i$. In contrast, if full correlation existed such that all the off-diagonal elements of $\boldsymbol R_i$ are equal to one,\footnote{Please note that such a case is only hypothetical and cannot exist in reality, as it indicates that all elements share the exact same physical location.} then $\text{tr}(\boldsymbol R_i^2) = N_i^2$.
	\begin{figure}[t]
		\centering
		{\includegraphics[width=4.9cm,height=4.9cm, trim={4.5cm 7.8cm 4.5cm 7.8cm},clip]{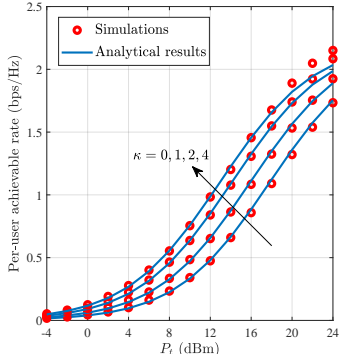}}
		\caption{Per-user achievable rate versus transmit power under different levels of VM noise concentration parameter $(\kappa)$ when $|\rho| = 0.8$, $N_1 = N_2 = 100$, and $M = 64$.}
		\label{Fig2}
	\end{figure}
	\section{Results and Discussion} \label{results}
	We start by defining the simulation setup. The BS is located at the origin of a $\mathrm {2D}$ plane such that $(x_{\text{BS}}, y_{\text{BS}}) = (0,0)$, while $(x_{I_1}, y_{I_1}) = (0,15)$, $(x_{I_2}, y_{I_2}) = (60,15)$, all in meters, and the $K$ users are located over a straight line between $(50, 0)$ and $(70,0)$, such that $(x_{U_k}, y_{U_k}) = (50 + \frac{k-1}{K-1}\times 20, 0), \ \forall k\in\mathcal U$. Unless stated otherwise, we also set $\lambda = 0.1\mathrm m$, $d_V = d_H = \lambda/4$, $\alpha = 2.7$, RIS element spacing $\varepsilon = \lambda/4$, $\sigma^2 = -94 \ \mathrm{dBm}$, and $K=4$. Finally, equal power allocation has been adopted among all users, and optimal RBD was utilized such that $\boldsymbol \Theta_i = \text{diag}\{\boldsymbol 1_{N_i}\}$.  
	\par Fig. \ref{Fig2} demonstrates the average per-user achievable rate performance. As one would expect, a larger transmit power budget at the BS leads to a higher achievable rate. In addition, the results also highlight the effect of phase-noise on the performance. In particular, higher values of $\kappa$ means that the distribution of the VM phase-noise is more centered around its mean, thereby allowing better phase-adjustment capabilities at both RISs and thus higher achievable rates. It is also worth highlighting that the analytical results closely match the ones obtained via computer simulations, which validate our closed-form expressions and prove their accuracy. 
	\par
\begin{figure}[t]
		\centering
		{\includegraphics[width=4.9cm,height=4.9cm, trim={4.5cm 7.8cm 4.5cm 7.8cm},clip]{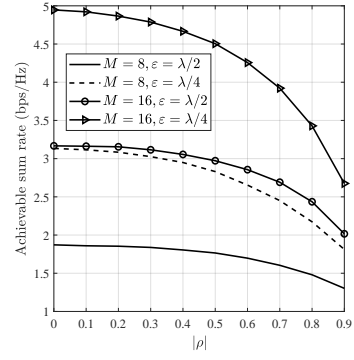}}
		\caption{Achievable sum rate versus $|\rho|$ for different number of BS antennas $(M)$ when $N_1=N_2 = 100$, $P_t = 20 \ \mathrm{dBm}$, and $\kappa = 4$.}
		\label{Fig3}
	\end{figure}

 Fig. \ref{Fig3} illustrates the effects of spatial correlation at both RISs as well as BS. The presented results firmly confirm our statements regarding the opposite roles of spatial correlation at BS and RISs for the considered MU-MISO network scenario, as indicated in both (\ref{lower_sinr}) and (\ref{eta_k_optimal}). More specifically, the results in Fig. \ref{Fig3} clearly show that higher spatial correlation at BS is not desired, while the opposite holds for the correlation at RISs (note that smaller element spacing $\epsilon$ means higher correlation at RIS). These observations indicate that for practical implementations, the miniaturization of RIS is highly desirable, as it leads to higher achievable rates due to the increased spatial correlation. Finally, as expected and also indicated by (\ref{lower_sinr}), increasing the number of BS antennas $M$ can largely enhance the achievable rate performance. 
	\section{Conclusion}\label{conclusions}
	We investigated the performance of a phase-noise impaired double-RIS MU-MISO network over spatially correlated channels. Closed-form expressions were derived for both the downlink achievable rate and optimal reflect beamforming design under MRT precoding scheme. Our results provided insight into the effects of phase-noise and spatial correlation at BS and RISs. In particular, it was observed that high spatial correlation is preferred at RISs, but not the BS, to achieve higher rates. Numerical results closely matched the analytical expressions which proved the validity and accuracy of our analysis. 
	\section*{Appendix}
	We start by introducing the lower-bound expression of the ergodic SINR, which relies on \textit{channel hardening}. In particular, considering the $k$th user, we have \cite{bjornson2017massive}
	{ \begin{align} \label{sinr_k}
			\underline{\mathrm{SINR}}_k = & \frac{p_k\left | \mathbb E\left\{\boldsymbol h_k^T\boldsymbol w_k\right\} \right |^2}{\sum\limits_{l\in\mathcal U} p_l \mathbb E\left\{\left | \boldsymbol h_k^T \boldsymbol w_l \right |^2\right\} - p_k\left | \mathbb E\left\{\boldsymbol h_k^T\boldsymbol w_k\right\} \right |^2 + \mathrm {Var} \left\{z_k\right\}}\nonumber \\ \stackrel {\mathrm{a_0}}{=} &  \frac{p_k \text{tr} \left(\boldsymbol \Psi_k\right)}{\sum\limits_{\substack{l\in \mathcal U}} p_l \frac{\text{tr} \left(\boldsymbol \Psi_k \boldsymbol \Psi_l\right)}{\text{tr} \left(\boldsymbol \Psi_l\right)}+\sigma^2},
	\end{align}}where $\boldsymbol \Psi_k$ is the channel covariance matrix for $k$th user, and the equality $\mathrm{(a_0)}$ holds only for the adopted MRT beamforming. Note that according to the central limit theorem, the cascaded channel admits to complex Gaussian distribution when $N_1$ and $N_2$ are large, and thus Lemma B.14 in \cite{bjornson2017massive} can be applied to obtain the expression in (\ref{sinr_k}). 
\par Next, we derive the channel covariance matrix of $k$th user. In particular, we have
	{\small \begin{align}\label{norm}
			& \hspace{-.3cm} \boldsymbol \Psi_k = \mathbb E\left\{\boldsymbol h_k\boldsymbol h_k^H\right\} \nonumber \\ & \hspace{0.15cm} = \mathbb E \Big\{ \boldsymbol{H}_{B1} \tilde{\boldsymbol{\Theta}}_{1}\boldsymbol{G} \tilde{\boldsymbol{\Theta}}_{2}\boldsymbol{q}_{2k}\boldsymbol{q}_{2k}^H \tilde{\boldsymbol{\Theta}}_{2}^H \boldsymbol{G}^H \tilde{\boldsymbol{\Theta}}_{1}^H \boldsymbol{H}_{B1}^H
			\nonumber \\ & \hspace{1cm} + \boldsymbol{H}_{B1} \tilde{\boldsymbol{\Theta}}_{1} \boldsymbol{q}_{1k}\boldsymbol{q}_{1k}^H \tilde{\boldsymbol{\Theta}}_{1}^H   \boldsymbol{H}_{B1}^H + \boldsymbol{H}_{B2} \tilde{\boldsymbol{\Theta}}_{2} \boldsymbol{q}_{2k} \boldsymbol{q}_{2k}^H \tilde{\boldsymbol{\Theta}}_{2}^H \boldsymbol{H}_{B2}^H
			\nonumber \\ & \hspace{1cm} + 2\Re \left\{\boldsymbol{H}_{B1} \tilde{\boldsymbol{\Theta}}_{1} \boldsymbol{q}_{1k}\boldsymbol{q}_{2k}^H \tilde{\boldsymbol{\Theta}}_{2}^H \boldsymbol{G}^H \tilde{\boldsymbol{\Theta}}_{1}^H \boldsymbol{H}_{B1}^H\right\} \nonumber \\ & \hspace{1cm} +  2\Re\left\{\boldsymbol{H}_{B2} \tilde{\boldsymbol{\Theta}}_{2} \boldsymbol{q}_{2k}\boldsymbol{q}_{2k}^H \tilde{\boldsymbol{\Theta}}_{2}^H \boldsymbol{G}^H \tilde{\boldsymbol{\Theta}}_{1}^H \boldsymbol{H}_{B1}^H\right\} \nonumber \\ & \hspace{1cm} + 2\Re\left\{\boldsymbol{H}_{B2} \tilde{\boldsymbol{\Theta}}_{2} \boldsymbol{q}_{2k}\boldsymbol{q}_{1k}^H \tilde{\boldsymbol{\Theta}}_{1}^H  \boldsymbol{H}_{B1}^H   \right\}\Big\}.
	\end{align}}In the following, we evaluate the expectation of each term in (\ref{norm}) separately. Specifically, for the first term, we have
	{\small \begin{align}\label{1st} 
			& \mathbb E \Big\{ \boldsymbol{H}_{B1} \tilde{\boldsymbol{\Theta}}_{1}\boldsymbol{G} \tilde{\boldsymbol{\Theta}}_{2}\boldsymbol{q}_{2k}\boldsymbol{q}_{2k}^H \tilde{\boldsymbol{\Theta}}_{2}^H \boldsymbol{G}^H \tilde{\boldsymbol{\Theta}}_{1}^H \boldsymbol{H}_{B1}^H \Big\}\nonumber \\ & = \beta_{2k} \mathbb E\left\{\boldsymbol{H}_{B1} \tilde{\boldsymbol{\Theta}}_{1}\boldsymbol{G}\tilde{\boldsymbol{\Theta}}_{2}\boldsymbol R_2^{\frac{1}{2}}\tilde{\boldsymbol{q}}_{2k}\tilde{\boldsymbol{q}}_{2k}^H \boldsymbol R_2^{\frac{1}{2}} \tilde{\boldsymbol{\Theta}}_{2}^H \boldsymbol{G}^H \tilde{\boldsymbol{\Theta}}_{1}^H \boldsymbol{H}_{B1}^H \right\}\nonumber \\ & = \beta_{2k} \mathbb E\left\{\boldsymbol{H}_{B1}\tilde{\boldsymbol{\Theta}}_{1}\boldsymbol{G} \bar{\boldsymbol{\Theta}}_{2} \boldsymbol \Theta_2 \boldsymbol R_2 {\boldsymbol{\Theta}}_{2}^H \bar{\boldsymbol{\Theta}}_{2}^H \boldsymbol{G}^H \tilde{\boldsymbol{\Theta}}_{1}^H \boldsymbol{H}_{B1}^H \right\}\nonumber \\ & \stackrel{\mathrm{a_1}}{=} \beta_{2k} \mathbb E\left\{\boldsymbol{H}_{B1}\tilde{\boldsymbol{\Theta}}_{1}\boldsymbol{G} \Big( \varphi^2\bar{\boldsymbol R}_2 + (1-\varphi^2)\boldsymbol I_{N_2}\Big) \boldsymbol{G}^H \tilde{\boldsymbol{\Theta}}_{1}^H \boldsymbol{H}_{B1}^H \right\}\nonumber \\ & = \beta_{2k} \Big(\varphi^2 \boldsymbol A + (1-\varphi^2)\boldsymbol B\Big),
	\end{align}}where $\bar{\boldsymbol R}_2 = \boldsymbol \Theta_2 \boldsymbol R_2 {\boldsymbol{\Theta}}_{2}^H$, and equality $\mathrm{(a_1)}$ holds from \cite[Eq.(13)]{papa1}. Furthermore, we have  
	{\small \begin{align} \label{A}
			&\hspace{-.25cm} \boldsymbol A = \mathbb E \Big\{ \boldsymbol{H}_{B1}\tilde{\boldsymbol{\Theta}}_{1}\boldsymbol{G}  \bar{\boldsymbol R}_2 \boldsymbol{G}^H \tilde{\boldsymbol{\Theta}}_{1}^H \boldsymbol{H}_{B1}^H \Big\}\nonumber \\ & = \beta_{G} \mathbb E \Big\{ \boldsymbol{H}_{B1}\tilde{\boldsymbol{\Theta}}_{1} \boldsymbol R_1^{\frac{1}{2}} \tilde{\boldsymbol{G}} \boldsymbol R_2^{\frac{1}{2}} \bar{\boldsymbol R}_2 \boldsymbol {R}_2^{\frac{1}{2}} \tilde{\boldsymbol{G}}^H \boldsymbol R_1^{\frac{1}{2}} \tilde{\boldsymbol{\Theta}}_{1}^H \boldsymbol{H}_{B1}^H \Big\}\nonumber \\ & = \beta_{G} \text{tr}\left(\boldsymbol R_2 \bar{\boldsymbol R}_2\right) \mathbb E \Big\{ \boldsymbol{H}_{B1}\bar{\boldsymbol{\Theta}}_{1} {\boldsymbol{\Theta}}_{1} \boldsymbol R_1 {\boldsymbol{\Theta}}_{1}^H \bar{\boldsymbol{\Theta}}_{1}^H \boldsymbol{H}_{B1}^H \Big\}\nonumber \\ & = \beta_{G} \text{tr}\left(\boldsymbol R_2 \bar{\boldsymbol R}_2\right) \mathbb E \Big\{ \boldsymbol{H}_{B1}\Big(\varphi^2 \bar{\boldsymbol R}_1 + (1-\varphi^2)\boldsymbol I_{N_1}\Big) \boldsymbol{H}_{B1}^H \Big\} \nonumber \\ & = \beta_{G} \text{tr}\left(\boldsymbol R_2 \bar{\boldsymbol R}_2\right) \Big(\varphi^2\mathbb E \Big\{ \boldsymbol{H}_{B1} \bar{\boldsymbol R}_1 \boldsymbol{H}_{B1}^H \Big\} \nonumber \\ & \hspace{2.5cm}+ (1-\varphi^2)\mathbb E \Big\{ \boldsymbol{H}_{B1} \boldsymbol{H}_{B1}^H \Big\}\Big) \nonumber \\ & = \beta_{G} \text{tr}\left(\boldsymbol R_2 \bar{\boldsymbol R}_2\right) \Big(\varphi^2\beta_{B1}\mathbb E \Big\{ \boldsymbol R_{B}^{\frac{1}{2}} \tilde{\boldsymbol H}_{B1} \boldsymbol R_1^{\frac{1}{2}} \bar{\boldsymbol R}_1 \boldsymbol R_1^{\frac{1}{2}} \tilde{\boldsymbol H}_{B1}^H  \boldsymbol R_{B}^{\frac{1}{2}}  \Big\} \nonumber \\ & \hspace{2cm} + (1-\varphi^2)\beta_{B1}\mathbb E \Big\{ \boldsymbol R_{B}^{\frac{1}{2}} \tilde{\boldsymbol H}_{B1} \boldsymbol R_1 \tilde{\boldsymbol H}_{B1}^H  \boldsymbol R_{B}^{\frac{1}{2}}  \Big\}\Big)\nonumber \\ & = \beta_{B1} \beta_{G} \text{tr}\left(\boldsymbol R_2 \bar{\boldsymbol R}_2\right) \Big(\varphi^2 \text{tr}\left(\boldsymbol R_1 \bar{\boldsymbol R}_1\right) + (1-\varphi^2) \text{tr}\left(\boldsymbol R_1\right)\Big) \boldsymbol R_B,
	\end{align}}where $\bar{\boldsymbol R}_1 = \boldsymbol \Theta_1 \boldsymbol R_1 {\boldsymbol{\Theta}}_{1}^H$, and the properties $\text{tr}\left(\boldsymbol XY\right) = \text{tr}\left(\boldsymbol YX\right)$, and $\mathbb E\{\boldsymbol V \boldsymbol U\boldsymbol V^H\} = \text{tr}\left(\boldsymbol U \right)\boldsymbol I_{d_1}$ were applied. The latter property holds for any  matrix $\boldsymbol V\in \mathbb C^{d_1\times d_2}$ with i.i.d entries of zero mean and unit variance, and $\boldsymbol U$ being a deterministic square matrix  with per-dimension size of $d_2$.  Following similar steps, the value of $\boldsymbol B$ can be given as
	{\small \begin{align} \label{B}
			& \boldsymbol B = \mathbb E\left\{\boldsymbol{H}_{B1}\tilde{\boldsymbol{\Theta}}_{1} \boldsymbol{G}  \boldsymbol{G}^H \tilde{\boldsymbol{\Theta}}_{1}^H \boldsymbol{H}_{B1}^H \right\}\nonumber \\ &  \hspace{0.25cm} = \beta_{B1} \beta_{G} \text{tr}\left(\boldsymbol R_2\right) \Big(\varphi^2 \text{tr}\left(\boldsymbol R_1 \bar{\boldsymbol R}_1\right) + (1-\varphi^2) \text{tr}\left(\boldsymbol R_1\right)\Big) \boldsymbol R_B.
	\end{align}}Furthermore, for the second term in (\ref{norm}), we have
	{\small \begin{align}\label{2nd} 
			& \mathbb E \Big\{ \boldsymbol{H}_{B1} \tilde{\boldsymbol{\Theta}}_{1} \boldsymbol{q}_{1k}\boldsymbol{q}_{1k}^H \tilde{\boldsymbol{\Theta}}_{1}^H   \boldsymbol{H}_{B1}^H \Big\} \nonumber \\ & = \beta_{B1} \beta_{1k} \Big(\varphi^2\text{tr} \left(\boldsymbol R_{1} \bar{\boldsymbol R}_1 \right) + (1-\varphi^2) \text{tr} \big(\boldsymbol R_1\big)\Big) \boldsymbol R_B.
	\end{align}}Similarly, the third term is given as
	{\small \begin{align} \label{3rd} 
			& \mathbb E \Big\{ \boldsymbol{H}_{B2} \tilde{\boldsymbol{\Theta}}_{2} \boldsymbol{q}_{2k}\boldsymbol{q}_{2k}^H \tilde{\boldsymbol{\Theta}}_{2}^H   \boldsymbol{H}_{B2}^H \Big\} \nonumber \\ & = \beta_{B2} \beta_{2k} \Big(\varphi^2\text{tr} \left(\boldsymbol R_{2} \bar{\boldsymbol R}_2 \right) + (1-\varphi^2) \text{tr} \big(\boldsymbol R_2\big)\Big) \boldsymbol R_B.
	\end{align}}In contrast, the expected value for each of the last three terms in (\ref{norm}) is equal to zero due to the independence between $\boldsymbol q_{1k}$ and $\boldsymbol q_{2k}$, as well as between $\boldsymbol H_{B1}$ and $\boldsymbol H_{B2}$. \par Next, we recall that $\text{tr}\left(\boldsymbol R_i\right) = N_i$, and after re-arranging the expressions and substituting the results of (\ref{A}) and (\ref{B}) into (\ref{1st}), followed by adding the results of (\ref{1st}), (\ref{2nd}) and (\ref{3rd}), we obtain
	\small {\begin{align} \label{final_term}
			&\boldsymbol \Psi_k  = \Big[\beta_{B1}\beta_{2k} \beta_G \Big(\text{tr}\left(\boldsymbol R_2 \bar{\boldsymbol R}_2\right) \Big( \varphi^4 \text{tr}\left(\boldsymbol R_1 \bar{\boldsymbol R}_1\right) + (\varphi^2-\varphi^4)N_1 \Big) \nonumber \\ & \hspace{1cm} + N_2\Big((\varphi^2-\varphi^4)\text{tr}\left(\boldsymbol R_1 \bar{\boldsymbol R}_1\right) + (1-\varphi^2)^2N_1\Big)\Big)\nonumber \\ & \hspace{1cm} + \beta_{B1}\beta_{1k} \Big(\varphi^2\text{tr}\left(\boldsymbol R_1 \bar{\boldsymbol R}_1\right) + (1-\varphi^2)N_1\Big) \nonumber \\ & \hspace{1cm} + \beta_{B2}\beta_{2k} \Big(\varphi^2\text{tr}\left(\boldsymbol R_2 \bar{\boldsymbol R}_2\right) + (1-\varphi^2)N_2\Big) \Big]\boldsymbol R_B \nonumber \\ & \hspace{0.5cm} = \eta_k \boldsymbol R_B.
	\end{align}}Finally, substituting the expression of $\boldsymbol \Psi_j\ (j\in \{k,l\})$ into (\ref{sinr_k}), and noting that $\text{tr}\left(\boldsymbol R_B\right) = M$, the expression in (\ref{lower_sinr}) is obtained after straightforward mathematical manipulations. 
	\bibliographystyle{IEEEtran}
	\bibliography{Double_RIS}
\end{document}